\documentclass[twocolumn,english,showpacs,superscriptaddress,prl]{revtex4}
\usepackage[T1]{fontenc}
\usepackage[latin9]{inputenc}
\usepackage{amsmath}
\usepackage{amssymb}
\usepackage{graphicx}
\usepackage{esint}

\usepackage{babel}

\begin{document}

\title{Mean Field Theory of Echo State Networks. }

\author{Marc Massar}

\affiliation{370 Central Park West, Appt. 511, NY 10025 New York, USA}

\author{Serge Massar}

\affiliation{Laboratoire d'Information Quantique, CP 225, Universit{\'e} libre
de Bruxelles (U.L.B.), Av. F. D. Roosevelt 50, B-1050 Bruxelles, Belgium}

\begin{abstract}
Dynamical systems driven by strong external signals are ubiquitous
in nature and engineering. Here we study \textquotedbl{}echo state
networks\textquotedbl{}, networks of a large number of randomly connected
nodes, which represent a simple model of a neural network, and have
important applications in machine learning. We develop a mean field
theory of echo state networks. The dynamics of the network is captured
by the evolution law, similar to a logistic map, for a single collective
variable. When the network is driven by many independent external
signals, this collective variable reaches a steady state. But when
the network is driven by a single external signal, the collective
variable is non stationary but can be characterized by its time averaged
distribution. The predictions of the mean field theory, 
including the value of the largest Lyapunov exponent, are compared
with the numerical integration of the equations of motion. 
\end{abstract}

\date{October 2012}

\maketitle
\section{Introduction.} 
Our understanding of non linear dynamical systems
and networks has made tremendous progress during the past decades.
In most cases the autonomous dynamics is studied. The situation where
the network is strongly driven by an external signal has so far been
less investigated even though it arises in many different contexts
in the natural and artificial world. Examples include networks of
interacting chemicals (proteins, RNA) in a cell driven by unpredictable
external chemical signals; networks of neurons driven by an external
sensory input; artificial neural networks and their applications in
machine learning; the response of population dynamics and ecological
networks to changes in external conditions such as the weather; the
responses of stock prices to economically significant news such as
a company earnings, or unemployment numbers. In all these cases taking
into account the external input is essential if one wants to understand
correctly the dynamics, both because the external input is often large
(it cannot be treated as a small perturbation), and because in some
cases the systems itself has been selected according to its response
to the fluctuating and unpredictable external variables.

The aim of the present work is to show, through the study of a specific
but important example, how mean field techniques can provide a detailed
understanding of dynamical networks strongly driven by an external
signal. In the mean field approach the average feedback of the variables
on themselves is taken into account through a self consistent equation,
while the correlations between individual variables are neglected.
The apparently extremely complicated dynamics of the network is thus
reduced to much simpler evolution equations for a few collective variables.
Previous applications of the mean field approach to dynamical systems
(but without including an external input), and in particular neural
networks, include e.g. \cite{key-AmitTsodyks91,key-Treves93,key-MattiaDelGiudice02,key-Sompolinsky}.
For previous studies of dynamical systems in the presence of external
signals (with however a quite different emphasis than in the present
work) see e.g. \cite{key-Lindnera04,key-Sagues07}. Mean field analysis of multi-population neural networks in the presence of stochastic noise have been recently presented in \cite{FaugerasTouboulCessac09,TouboulHermannFaugeras12}.

The specific system we will consider is taken from the field of artificial
neural networks. It consists of a network of randomly connected idealized
neurons evolving in discrete time, and driven by an external time
dependent signal, known in the machine learning community as an ``echo
state network'' \cite{key-Jaeger-echostate-report,key-Jaeger-Haas-Science},
see also the continuous time analog with no input studied in \cite{key-Sompolinsky}.
Such systems, when supplemented by a single linear output layer, fall
within the class of ``reservoir computers'' \cite{key-Jaeger-echostate-report,key-MaasNatschlagerMarkram02,key-Jaeger-Haas-Science,key-Verstraeten-Schrauwen-unification-07}
and currently hold records for several highly non trivial machine
learning tasks such as time series prediction or some speech recognition
benchmarks, see e.g. \cite{key-LukoseviciusJaeger-review} for a review.
Because of their importance in the machine learning community, it
is highly desirable to better understand the dynamics of these systems.
In addition they can serve as toy models for investigating the dynamics
of neural networks, and more generally any recurrent dynamical systems,
driven by external inputs.

Here we show that the dynamics of echo state networks can be concisely described by a single collective variable, namely the variance $\sigma^{2}(t)$ of the variables describing the echo state network. In the limit when the number of internal variables is large (which is the case in reservoir computing applications), the variance obeys a closed evolution equation, similar to the logistic map, but with a source term  (due to the source term that drives the echo state network). We further show how to derive the onset of chaos in echo state networks, and we derive the Lyapunov exponent. In the case of a sigmoid non linearity it can be shown that the external input stabilizes the system, as is well known in the community working on echo state networks (see e.g. \cite{luk}). We note that Lyapunov exponents for dynamical systems driven by stochastic noise were studied in e.g. \cite{Pikovsky92} where it was shown that the noise can dramatically change the stability of the system. Stabilization of chaotic systems by controlled inputs was described  in \cite{OttGrebogiYorke90}. 
Throughout our work we compare in the figures the predictions of the mean field theory with the exact integration of the equations of motion. In all cases we find excellent agreement.

\section{Echo State Networks}

An echo state network consists of a large number $N$ of artificial
neurons evolving in discrete time $t\in\mathbb{Z}$. We denote by
$a_{i}(t)$ the ``activation potential'' of neuron $i$ at time
$t$. At time $t+1$, neuron $i$ sends a signal to the other neurons
with strength $x_{i}(t+1)$ given by 
\begin{equation}
x_{i}(t+1)=f\left(a_{i}(t)\right)\ ,\label{eq:xRC}
\end{equation}
where the function $f$ is taken to be a sigmoidal functions, i.e.
$f$ is odd, monotonously increasing, has finite limit for large $a$,
and its first derivative $f'(a)$ decreases monotonously for positive
$a$. By rescaling $x$ and $a$ we can redefine $f(a)\to\alpha f(\beta a)$.
We choose the scales such that $f'(0)=1$ and $\lim_{a\to\infty}f(a)=1$.
In the illustrative figures, we choose for $f$ the hyperbolic tangent
$f\left(a\right)=\tanh\left(a\right)$, as this is the form most often used in echo state networks.

The update rule for the activation potentials is 
\begin{equation}
a_{i}(t)=\sum_{j=1}^{N}w_{ij}x_{j}(t)+u_{i}s(t)\ ,\label{eq:aRC}
\end{equation}
where $w_{ij}$ is a time independent coupling matrix which gives
the strength of the coupling of neuron $j$ to neuron $i$, $s(t)$
is the time dependent external input, and $u_{i}$ is a time independent
vector which determines the strength with which the input is coupled
to neuron $i$.

In echo state networks, the $w_{ij}$ and $u_{i}$ are chosen independently
at random, except for global scaling factors $w_{ij}\to\mu w_{ij}$,
$u_{i}\to\nu u_{i}$. By adjusting these scaling factors and by
using an optimized linear readout it is possible to obtain excellent
performance on a variety of machine learning tasks. The general heuristic
is that the factor $\alpha$ should be adjusted for the system to
be at the threshold of chaos, whereupon the response of the neural
network to the input is highly complex, but deterministic. Previous
studies of the dynamics of echo state networks have aimed to understand
how different dynamical regimes are related to changes in their information
processing capability \cite{key-SchrauwenBusingLegenstein2009}. Most
closely connected to the present work is \cite{key-HermanSchrauwen-infiniteEchoStateNetworks}
where, based on earlier work on feed forward networks \cite{key-Williams-inifiniteneuralnetworks},
the information processing capability of echo state networks could
be studied in the limit where the number $N$ of neurons tends to
infinity.

\section{Variations around the Echo State Network equations.}

The echo state network equations (\ref{eq:xRC}) and (\ref{eq:aRC}) can be modified in a number of ways, all of which are also amenable to treatment using the mean field approximation. Comparison between these different dynamical systems helps understand the generality but also limitations of the mean field approximation. We list here the most important such generalizations. In all cases eq. (\ref{eq:xRC}) is unchanged. It is the update rule for the activation potential eq. (\ref{eq:aRC}) that is modified.

\subsection{Multiple inputs.}

In some applications of echo state networks, such as e.g. preprocessed speech signals or image processing, 
there are multiple inputs $s_1(t),\ldots , s_K(t)$ that drive the system. This is also important in e.g. spatiotemporal processing by the cortex\cite{BuonomanoMaass}.
We model the multiple inputs $s_l$ as independent random variables.

In this case the update rule for the activation potential becomes:
\begin{equation}
a_{i}(t)=\sum_{j=1}^{N}w_{ij}x_{j}(t)+\sum_{l=1}^K u_{il}s_l(t)\ ,\label{eq:aRC_MultIn}
\end{equation}
where $w_{ij}$ and $u_{il}$ are time independent coefficients. 

\subsection{Independent inputs}
When $K\to \infty$ then the source terms in eq. (\ref{eq:aRC_MultIn}) become independent random variables (provided the $u_{il}$ are rescaled to keep the variance of the source term finite), and  the update rule for the activation potential becomes:
\begin{equation}
a_{i}(t)=\sum_{j=1}^{N}w_{ij}x_{j}(t)+s_i(t)\ ,\label{eq:aRC_IndepIn}
\end{equation}
where $w_{ij}$ are time independent coefficients, and the $s_i$ are independent random variables.

\subsection{Annealing approximation}
In the annealing approximation (see e.g. \cite{DerridaPomeau86,Bertschinger04} in the case of discrete variables), the update rule for the activation potential is:
\begin{equation}
a_{i}(t)=\sum_{j=1}^{N}w_{ij}(t)x_{j}(t)+u_{i}(t)s_(t)\ ,\label{eq:aRC_Anneal}
\end{equation}
where $w_{ij}(t)$ and $u_{il}(t)$ are time dependent coefficients that, at each time $t$, are drawn independently at random from the same distribution.
In the annealing approximation any structure arising from the fixed values of the coefficients $w_{ij}$ and $u_i$ is erased, since these coefficients change at each time $t$. The annealing approximation can be equally applied to the case of multiple inputs eq. (\ref{eq:aRC_MultIn}) and independent inputs eq.(\ref{eq:aRC_IndepIn}).

The mean field approach (discussed in the remainder of this article) can be equally applied in the case of the annealing approximation. The resulting equations are identical to those obtained from equations (\ref{eq:xRC}, \ref{eq:aRC}, \ref{eq:aRC_MultIn}, \ref{eq:aRC_IndepIn}). That is the mean field approach cannot reveal structure that arises from the fixed values of the coefficients $w_{ij}$ and $u_{il}$.

\section{Mean Field approximation.}

\subsection{Mean field equations.}\label{MFE}

\begin{figure}
\includegraphics[scale=0.4]{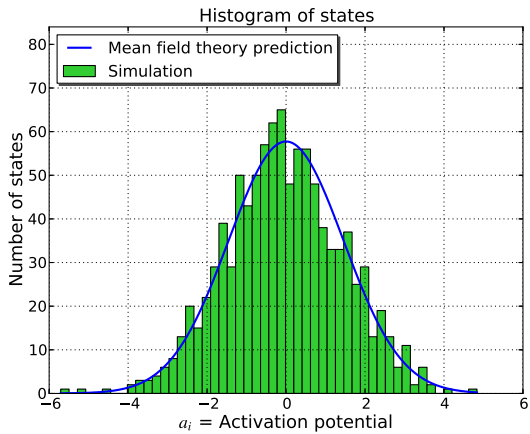} \caption{Distribution of the activation potential $a_{i}$. A reservoir with
size $N=1000$ and normalized gain of $g=2$ and no source was run for
200 time steps, then the histogram of the activation potential was
plotted in green. For comparison a Gaussian with the variance predicted
by the mean field theory was plotted in blue.\label{fig:hist_no_source}}
\end{figure}

 The key insight behind the present
work is to make the assumption that, at each time $t$, the $x_{i}(t)$
behave as independent identically distributed random variables which
are also independent of the $w_{ij}$ and the $u_{i}$. Then the term
$\sum_{j=1}^{N}w_{ij}x_{j}(t)$ in eq. (\ref{eq:aRC}) is a sum of
many identically distributed independent variables, and the law of
large numbers tells us that this sum is distributed as a Gaussian, see fig. \ref{fig:hist_no_source}.

With this assumption we can compute the distribution of $a_{i}(t)$,
and then using eq. (\ref{eq:xRC}) the distribution of $x_{i}(t+1)$.
This analysis will yield a very simple one dimensional recurrence,
similar to the logistic map, which captures the essence of the dynamics
of the echo state network.

In more detail, we reason as follows. Because the function $f$ is
odd, the distribution from which are drawn the $x_{i}(t)$ has mean
zero $\langle x_{i}(t)\rangle=0$, where $\langle\rangle$ denotes
ensemble average, i.e. average over the index $i$ at fixed time $t$.
We denote the variance of the $x_{i}(t)$ by $\langle x_{i}^{2}(t)\rangle=\sigma^{2}(t)$.
Assuming that the $w_{ij}$ are drawn independently at random from
a distribution with mean zero $E\left[w_{ij}\right]=0$ and variance
$E\left[w_{ij}^{2}\right]=w^{2}$, and introducing the rescaled gain
$g$ as 
\[
g^{2}=Nw^{2}
\]
we find that the term 
\[
\sum_{j=1}^{N}w_{ij}x_{j}(t)\sim N(0,g^{2}\sigma^{2}(t))
\]
has Gaussian distribution. We now assume for simplicity that the $u_{i}$
are drawn independently at random from a Gaussian distribution with
zero mean and variance $u^2$:
\[
u_i\sim N(0,u^2)\ .
\]
Then the activation potential
\begin{eqnarray}
a_{i}(t)&=&\sum_{j=1}^{N}w_{ij}x_{j}(t)+u_{i}s(t)
\nonumber\\
&\sim& N(0,g^{2}\sigma^{2}(t)+u^2s^{2}(t))
\nonumber\end{eqnarray}
also has a Gaussian distribution. (If the $u_{i}$ are drawn from
a distribution other than Gaussian, then the distribution of the $a_{i}(t)$
can in principle be calculated, but the expressions will be more
complicated). For future use we denote the variance of the $a_{i}(t)$
as 
\[
\Sigma^{2}(t)=g^{2}\sigma^{2}(t)+u^2s^2(t)\ .\] 
Finally, the distribution
of $x_{i}$ at time $t+1$ is given by 
\[x_{i}(t+1)\sim f(N(0,\Sigma^2(t)))\  .\]
We thus obtain a closed one-dimensional recurrence for the variances
of $x_{i}(t)$ and $a_{i}(t)$: 
\begin{eqnarray}
\Sigma^{2}(t) & = & g^{2}\sigma^{2}(t)+u^2s^{2}(t)\nonumber \\
\sigma^{2}(t+1) & = & F\left(\Sigma^{2}(t)\right)\label{eq:recurrence}
\end{eqnarray}
where 
\begin{eqnarray}
F\left(\Sigma^{2}\right) & = & \int da\ f^{2}(a)\ \frac{\exp\left[-\frac{a^{2}}{2\Sigma^{2}}\right]}{\sqrt{2\pi\Sigma^{2}}}\nonumber \\
 & = & \int dy\ f^{2}(\Sigma y)\ \frac{\exp\left[-\frac{y^{2}}{2}\right]}{\sqrt{2\pi}}\ .\label{eq:F(Sigma)}
\end{eqnarray}
From the properties of the sigmoid funtion $f$ (given below eq. (\ref{eq:xRC})), it follows that $F\left(\Sigma^{2}\right)$
satisfies: $F\left(0\right)=0$, $F\left(+\infty\right)=1$, $dF/d\Sigma^{2}>0$,
$dF/d\Sigma^{2}(\Sigma^{2}=0)=1$.

In the illustrative figures discussed below we take $f(a)=\tanh(a)$
as this is the case most used in applications of echo state networks.
The integral eq. (\ref{eq:F(Sigma)}) yielding the stationary solution $F$
must then be carried out numerically (this can be done very efficiently).

It is however interesting to note that the function $F(\Sigma^{2})$
can be computed analytically in two cases. The first does not
correspond to a sigmoid function $f$, but it is of interest as it
has been used in a recent experimental implementation of reservoir
computing \cite{key-Paquotetal-optoelectRC,key-Larger}:
if $f(a)=\sqrt{2}\sin(\frac{a}{\sqrt{2}})$, then $F(\Sigma^{2})=1-\exp\left(\Sigma^{2}\right)$.
The second case (obtained by first evaluating $\frac{dF(\Sigma^2) }{d\Sigma^2}$, see \cite{key-Williams-inifiniteneuralnetworks}) corresponds to a sigmoid function $f$: 
if $f(a)=\mathrm{erf}(\frac{\sqrt{\pi}}{\sqrt{2}}a)$, then $F(\Sigma^{2})=-1+\frac{4}{\pi}\arctan\left(\sqrt{1+2\Sigma^{2}}\right)$. 
Results for these cases are similar to those when $f(a)=\tanh(a)$ and also show very good agreement between the mean field approximation and the integration of the exact equations (\ref{eq:xRC}, \ref{eq:aRC}). (Figures for these cases are not shown).

\subsection{Solution of the mean field equations.}

We now consider the solutions of the mean field equations, i.e. the coupled one dimensional recurrence eq. (\ref{eq:recurrence}).
We first consider the case when there is no source $s^{2}=0$. When
$g<1$, there is a single stationary solution to eq.  (\ref{eq:recurrence}):
$\sigma^{2}=\Sigma^{2}=0$,
corresponding to a quiescent system. $g=1$ corresponds to a branching
point. When $g>1$, the stable stationary solution
of eq.  (\ref{eq:recurrence})
 is different from
zero: $\sigma^{2}>0$, $\Sigma^{2}>0$. 
We will see below that when $g>1$, not only is $\Sigma^{2}>0$ but the Lyapunov exponent of the system eqs. (\ref{eq:xRC}, \ref{eq:aRC})  is greater than $1$. This therefore corresponds to a chaotic regime.
In the limit of infinite gain
$g\to\infty$, we have $\sigma^{2}\to1$, $\Sigma^{2}\to g^{2}$. 

When the source $s(t)$ is non-zero, integrating the recurrence eq. (\ref{eq:recurrence})
 yields a distribution of values
for $\sigma^{2}$ and $\Sigma^{2}$. In the figures, we take for illustrative
purposes the $s(t)$ to be independently drawn at each time $t$ from
the same probability distribution, that is we assume there are no
temporal correlations between successive values of the source. For
definiteness we take the source
\[
s(t)\sim N(0,s^{2})
\] 
to be distributed according
to a Gaussian with zero mean and variance $s^{2}$. 
We denote
\[
\xi^2=u^2s^2\ .
\]

\begin{figure}
\includegraphics[scale=0.4]{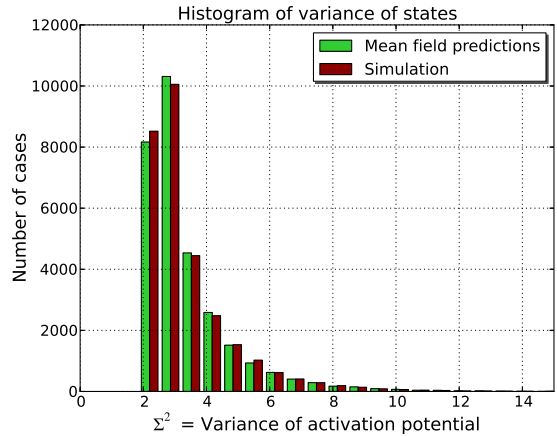} 
\caption{Histogram of the variance $\Sigma^{2}$ of the activation potential in the presence
of source. A reservoir of size $N=500$ with normalized gain $g=2$
and a single i.i.d normal source with variance $\xi^{2}=1$ was run for 200 time
steps to remove the influence of the initial state. Then the reservoir
was run for 30,000 time steps, the variance of the activation potential
$\Sigma^{2}(t)$ was collected, and its histogram plotted in red. The
same procedure was done following the mean field theory prediction
of eq. (\ref{eq:aRC}) and plotted in green.
\label{fig:Histogram-source}}
\end{figure}

Comparison of
the mean field theory and the exact distribution for $\sigma^{2}(t)$
obtained by integrating the equations of motion is given in fig.
\ref{fig:Histogram-source}.
And in fig. \ref{fig:Std_Var_vs_SourceVol} we find excellent agreement between the mean field theory predictions and the exact solutions for
the mean (over time) of $\Sigma^{2}(t)$
 for input strength $\xi^2=0.2$ as a function of $g^2$.

\begin{figure}
\includegraphics[scale=0.4]{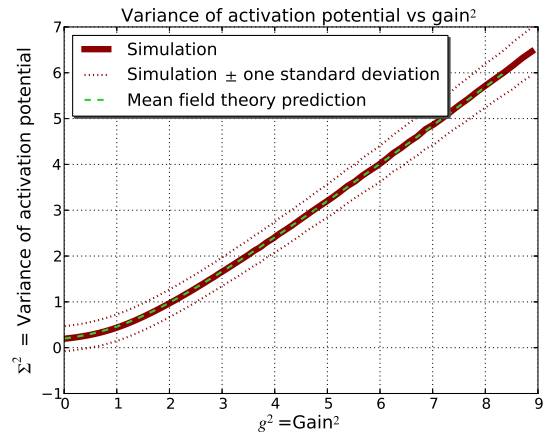} \caption{Variance of the activation potential $\Sigma^{2}$ as a function of
normalized gain $g$. For each activation potential, a reservoir of
size $N=500$ with a single i.i.d. Gaussian source with variance $\xi^{2}=0.2$
was run according to eqs. (\ref{eq:xRC}, \ref{eq:aRC}) for 200 time steps
to eliminate transitory effects, then the variance $\Sigma^{2}(t)$ of
the activation potential was recorded for 2000 time steps.
For each gain, the mean and standard deviation (over time) of $\Sigma^{2}(t)$
was computed. The mean is plotted in red, and mean $\pm$ one standard
deviation are plotted in doted red. The green dashed curve (superimposed on the red curve in the figure) corresponds
to the predictions of the mean field theory computed according to eq. (\ref{eq:recurrence}). The asymptotic behaviors discussed in section \ref {MFMI}, $\lim_{g^2\to 0}\Sigma^{2}=\xi^{2}=0.2$ and $\lim_{g^2\to \infty}\Sigma^{2}=g^2$, are clearly visible on the graph.
\label{fig:Variance-vs-gain}}
\end{figure}

\subsection{Mean field equations in the case of multiple inputs.}\label{MFMI}

We now consider the mean field approximation in the case where there are multiple inputs eq. (\ref{eq:aRC_MultIn}) and independent inputs eq. (\ref{eq:aRC_IndepIn}). The new features that arise in this case can be understood qualitatively as follows. When there is a single input $s(t)$, then if at some time $t$, $s(t)$ is large (small), then the $a_i(t)$ will have larger (smaller) absolute value than its average, and $\Sigma^2(t)$ will be larger (smaller). Thus $\Sigma^2(t)$ fluctuates in time. When there are $K$ independent inputs, the same phenomenon occurs, but is less pronounced since the fluctuations of the different inputs $s_l(t)$ tend to counterbalance each other. In the limit when there are infinitely many inputs, or equivalently when all the inputs are independent, the time dependent fluctuations of the individual inputs completely average out when we compute $\Sigma^2(t)$, which becomes time independent.

To analyze the case of multiple inputs,
we take the coefficients $u_{il}$ to be independent random variables drawn from a normal distribution
\[
u_{il}\sim N(0,u^2) .
\] 
We take the source terms $s_l(t)$ to be independent random variables, with zero mean and variance
\[
s^2 = \mbox{var}\left[s_l(t)\right]\ .
\] 
We denote by 
\[
\xi^2=K u^2 s^2
\] 
the variance of the source term $\sum_{l=1}^K u_{il} s_l(t)$.

The reasoning leading to the mean field equations can then be followed exactly as in section \ref{MFE} and one finds the same equations:
\begin{eqnarray}
\Sigma^{2}(t) & = & g^{2}\sigma^{2}(t)+v^2_K(t)\nonumber \\
\sigma^{2}(t+1) & = & F\left(\Sigma^{2}(t)\right)\ .\label{eq:recurrenceMult}
\end{eqnarray}
The dependence on the number $K$ of inputs only appears in the source term $v_K(t)$ which is a sum
of $K$ squares of Gaussians. It is therefore distributed as a Chi-squared with $K$ degrees of freedom
\[
v^2_K(t)\sim u^2 s^2 \chi^2(K) .
\] 
with expectation and variance
\[
E \left[v^2_K(t)\right]=\xi^2 
\quad , \quad 
\mbox{var}\left[v^2_K(t)\right]=2 K u^4 s^4 = \frac{2 \xi^4 }{K}
\] 
(where the expectations are time averages).

Thus if we keep the strength of the source term constant, that is if we keep $\xi^2$ constant, but increase the number of source terms, then the variance of the time dependent source term $v^2_K(t)$ in the mean field eqs. (\ref{eq:recurrenceMult}) decreases. In the limit $K\to \infty$, the source term becomes time independent. This is the form of  the mean field equation that obtains in the case
of independent inputs eq. (\ref{eq:aRC_IndepIn}).

We now  discuss in a little more detail the later case of independent inputs eq. (\ref{eq:aRC_IndepIn}). This corresponds to a time independent source term in eq. (\ref{eq:recurrenceMult}): $v^2_K(t)=\xi^2$.
The recurrence eq. (\ref{eq:recurrenceMult}) then admits a stationary solution. When $\xi^{2}>0$, the
stationary solution of eqs. (\ref{eq:recurrenceMult}) is always different from zero: $\sigma^{2}>0$,
$\Sigma^{2}>0$. For very small gain $g\to 0$ we have $\Sigma^{2}=\xi^{2}$,
$\sigma^{2}=F(\xi^{2})$. For large gain $g\to \infty$ the stationary solution tends to the solution in the absence of source $\sigma^{2}\to1$, $\Sigma^{2}\to g^{2}$.

In fig. \ref{fig:Variance-vs-gain} we compute the variance $\Sigma^2$ of an echo state network driven by a single source when the source term is weak $\xi^2=0.2$. Because the source term is weak, the cases of a single input and of multiple inputs are similar: the time average of the variance is identical to the stationary solution, and the standard deviation of $\Sigma^2$ is small. The asymptotic behaviors 
$\lim_{g^2\to 0}\Sigma^{2}=\xi^{2}=0.2$ and $\lim_{g^2\to \infty}\Sigma^{2}=g^2$ are clearly visible on the graph.

\begin{figure}
\includegraphics[scale=0.4]{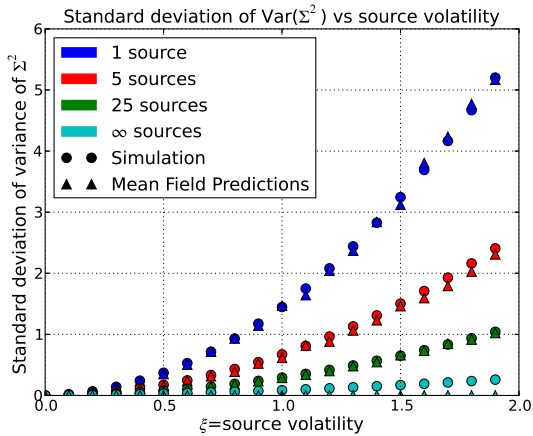} \caption{
Standard deviation of  $\Sigma^{2}(t)$ as a function of
source volatility $\xi$. The normalized gain was chosen to be $g=0.9$. Reservoirs of
size $N=500$ with $K=1, 5, 25, 500$ i.i.d. Gaussian source with total variance $\xi^{2}$
were simulated according to eqs. (\ref{eq:xRC}, \ref{eq:aRC_MultIn}). The reservoirs were run
 for 200 time steps to eliminate transitory effects, then the variance $\Sigma^{2}(t)$ of
the activation potential was recorded for 2000 time steps.
The standard deviation (over time) of $\Sigma^{2}(t)$
was computed and is plotted in the figure using dots. 
This is compared  the predictions of the mean field theory eq. (\ref{eq:recurrenceMult}), plotted using triangles.
Note that  the mean field predictions go to zero as $K \rightarrow \infty$, but we can not have more than 500
sources for a reservoir of this size. 
\label{fig:Std_Var_vs_SourceVol}}
\end{figure}

In fig. \ref{fig:Std_Var_vs_SourceVol} we compute the standard deviation of $\Sigma^2(t)$ of an echo state network driven by one source, by multiple source, and in the case of independent sources, as a function of the source volatility $\xi^2$. We compare the predictions of the mean field theory eq. (\ref{eq:recurrenceMult})and of the exact equations (\ref{eq:xRC},\ref{eq:aRC},\ref{eq:aRC_MultIn},\ref{eq:aRC_IndepIn}).

\section{Stability.} 
The mean field theory also allows a derivation
of the Lyapunov exponents of the echo state network. For simplicity we discuss only the case of a single input. 
Suppose that an echo state network is run until it reaches a typical state. At some time, say $t=0$ the solution is slightly perturbed, and then let to evolve. We therefore have two neighbouring solutions $a_i(t)$ and $a_i'(t)$. Denote by $\delta(t)= a_i(t) -a_i'(t)$. The largest  Lyapunov exponent of the system is given by $\Lambda=\lim_{t\to\infty, \delta(0)\to 0}\sqrt[t]{\delta^{2}(t)/\delta^{2}(0)}$.

\begin{figure}
\includegraphics[scale=0.4]{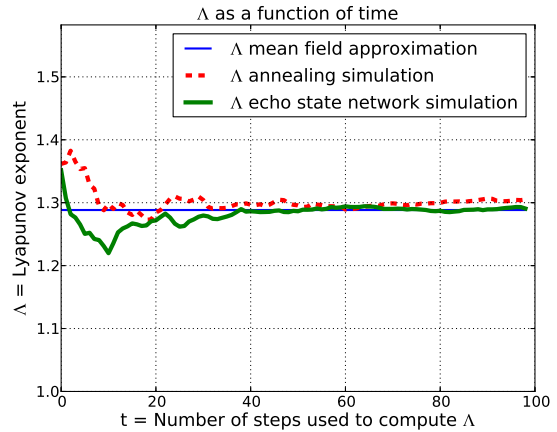} \caption{Convergence of Lyapunov exponent estimate. A
single reservoir of size $N=500$, normalized gain $g=2$, and source
variance $\xi^{2}=0.2$ was run for 200 steps. Then the states were
perturbed with i.i.d. Gaussians with standard deviation $10^{-10}$ and the size
of the perturbation was recorded for 100 time steps. The Lyapunov exponent
$\Lambda$ was computed for the different lags: $\sqrt[t]{\delta^{2}(t)/\delta^{2}(0)}$.
The green line is the result of a simulation done according to eqs. (\ref{eq:xRC}, \ref{eq:aRC}), the doted red line is the result of a simulation done in the annealed approximation, the straight blue line is the mean field theory prediction
computed from (\ref{eq:Stability-delta2}).}
\label{Lyapunov_n_steps}
\end{figure}

The mean field theory allows one to evaluate the largest Lyapunov exponent $\Lambda$. We suppose that the two neighboring
solutions $a_{i}(t)$ and $a_{i}'(t)$ have joint Gaussian distribution:
$\frac{a_{i}(t)+a_{i}'(t)}{2}\sim N(0,\Sigma^{2}(t))$, $a_{i}(t)-a_{i}'(t)\sim N(0,\delta^{2}(t))$.
We take $\delta^{2}(t)$ small and wish to compute how it evolves
with time. We have 
\begin{eqnarray}
&a_{i}(t+1)-a'_{i}(t+1)\quad \quad\quad \quad\quad&\nonumber\\
&\quad \quad=\sum_{j}w_{ij}f'\left(\frac{a_{i}(t)+a'_{i}(t)}{2}\right)\left(a_{i}(t)-a'_{i}(t)\right)&\nonumber\\
& \quad \quad \quad+O\left((a-a')^{2}\right)&\end{eqnarray}
where $f'=df(a)/da$ is the derivative of $f$ with respect to its argument.
From this equation we derive that 
\begin{eqnarray}
\delta^{2}(t+1) & = & \delta^{2}(t)\Lambda(\Sigma^{2}(t))+O\left(\delta^{3}(t)\right)\nonumber \\
\Lambda(\Sigma^{2}) & = & g^{2} \int da\ f'^{2}(a)\ \frac{\exp\left[-\frac{a^{2}}{2\Sigma^{2}}\right]}{\sqrt{2\pi\Sigma^{2}}}\ .\label{eq:Stability-delta2}
\end{eqnarray}
In fig. \ref{Lyapunov_n_steps}  we compare the Lyapunov exponents computed from the exact equations of motion and from the mean field theory.

From  expression (\ref{eq:Stability-delta2}) we deduce
two interesting properties.
First, in the absence of source the quiescent stationary solution
$\sigma^{2}(t)=\Sigma^{2}(t)=0$ has largest Lyapunov exponents smaller
than $1$ for $g<1$, and largest Lyapunov exponents larger than $1$
for $g>1$. (This follows from the fact that the integral in eq. (\ref{eq:Stability-delta2})
equals $1$ in the limit $\Sigma^{2}\to0$, since $f'(0)=1$). Thus,
in the absence of source, $g=1$ is the threshold for chaos in the
dynamical system.

\begin{figure}
\includegraphics[scale=0.4]{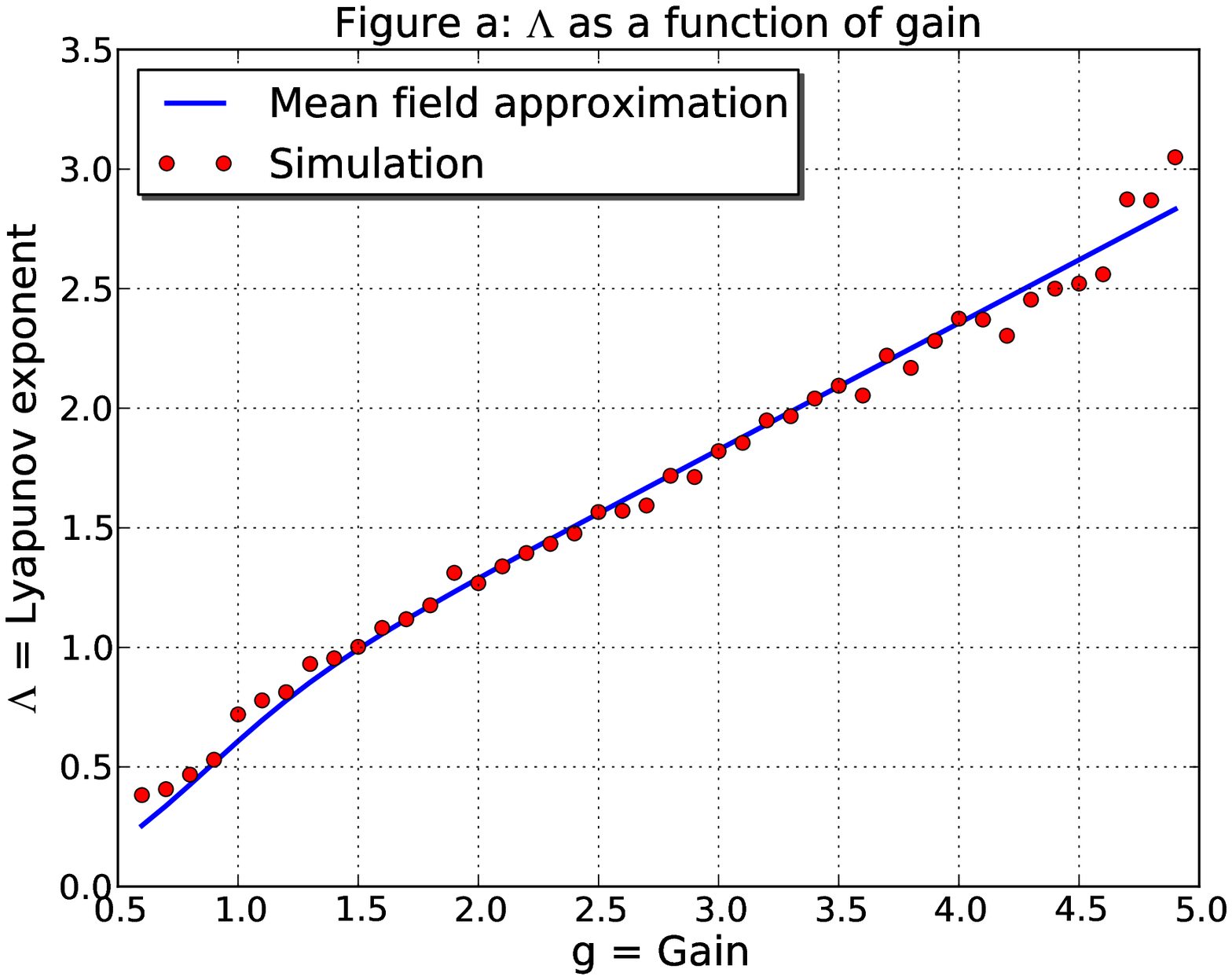} \includegraphics[scale=0.4]{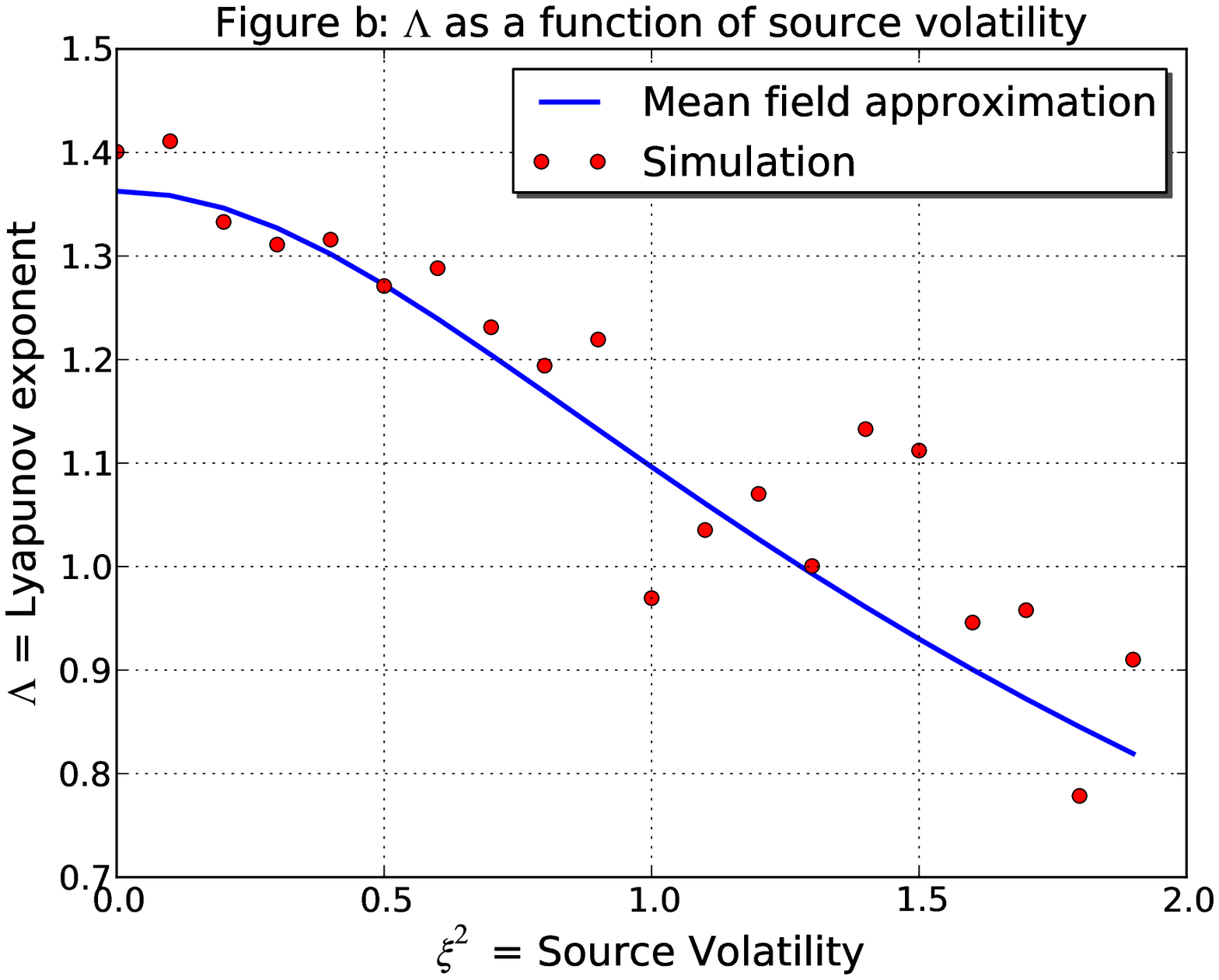}
\caption{Lyapunov exponent as a function of gain (panel a) and Lyapunov exponent as a function
of source volatility (panel b). An $N=500$, $g=2$ reservoir was run for 200 time
steps, than perturbed by i.i.d. Gaussians with standard deviation $10^{-12}$ and
run for 20 time steps. The Lyapunov exponent was computed
based on how much the perturbation changed, and plotted in red. The mean field theory prediction is in blue.
In panel a the source variance is $\xi^2=0.2$; in panel b the gain is
$g=2$. }
\label{Lyapunov_vs_gain_source_vol} 
\end{figure}

Second, for sigmoidal functions $f$, we find the property (well known
to those who use echo state networks for machine learning tasks, see e.g. \cite{luk}) that
increasing the strength of source term $\xi^{2}$ stabilizes the system.
Indeed when $\xi^{2}$ increase, $\Sigma^{2}$ also increases. For
sigmoidal functions $f'(a)$ is a decreasing function of $|a|$, and
hence the integral in eq. (\ref{eq:Stability-delta2}) decreases when
$\Sigma^{2}$ increases. See fig. \ref{Lyapunov_vs_gain_source_vol} for illustrations of this prediction.

\section{Conclusion.}
 In the present work we have shown that mean
field theory can be applied to large random networks driven by external
signals. The agreement, at least in the specific case of echo state networks, with the exact integration of the equations of motion is remarkably good, as illustrated by the figures.

The mean field theory itself is closely related to the 
annealed approximation wherein the $w_{ij}(t)$ and the $u_{i}(t)$
are redrawn independently at random at each time $t$ from the same
distribution, i.e. the coupling coefficients become time dependent, see  eq. (\ref{eq:aRC_Anneal}).
We expect the mean field theory to be exact in the large $N$ limit
of the annealed approximation.

Compared with the case when there is no external signal, we recover
a number of features which are well known empirically to people working
with echo state networks, but which have not been derived analytically
before. In particular we find that if in the absence of external signal
the system has a trivial stable state (corresponding in our analysis
to the case $g<1$), then in the presence of external signal the dynamics
becomes non trivial. We also find that the presence of the external
signal tends to stabilize the system (i.e. Lyapunov exponents which
decrease when the external signal increases).

We also compare the cases where the dynamical system is driven by
independent input signals, and by a single or a small number of input signals. We
find a qualitative difference. Namely in the first case the mean field
theory predicts that the collective variables take on stationary
values, whereas in the second case the mean field theory predicts
their statistical distribution.

The present work should provide the basis for 
further development of the theory of dynamical systems 
in the highly important case in practice 
when they are driven by time dependent external signals. 

{\bf Acknowledgments.} We acknowledge funding from the FRS-FNRS, the IAP under project Photonics@be, the Action de Recherche Concert{\'e}e.


\begin{thebibliography}{References}
\bibitem{key-AmitTsodyks91}D. J. Amit and M. Tsodyks, Network \textbf{2},
259, 1991

\bibitem{key-Treves93}A. Treves, Network \textbf{4}, 259, 1993

\bibitem{key-MattiaDelGiudice02}M. Mattia and P. Del Giudice, Phys.
Rev. E \textbf{66}, 051917, 2002

\bibitem{key-Sompolinsky}H. Sompolinsky, A. Crisanti, H. J. Sommers,
 Phys. Rev. Lett. \textbf{61}, pp. 259-262 (1988)

\bibitem{key-Lindnera04} B. Lindner, J. Garcia-Ojalvo, A. Neiman, 
L. Schimansky-Geier, 
Phys. Reports \textbf{392}, 321\textendash{}424 (2004)

\bibitem{key-Sagues07} F. Sagues, J. M. Sancho, J. Garcia-Ojalvo,
Rev. Mod. Phys. \textbf{79}, 829\textendash{}882 (2007)

\bibitem{FaugerasTouboulCessac09}
O. Faugeras, J. Touboul and B. Cessac, Front. Comput. Neurosci. {\bf 3}, pp. 1-28 (2009)

\bibitem{TouboulHermannFaugeras12}
J. Touboul, G. Hermann, O. Faugeras,
SIAM J. Appl. Dyn. Syst. {\bf 11}, pp. 49-81 (2012) 

\bibitem{key-Jaeger-echostate-report}H. Jaeger, 
Fraunhofer Institute for Autonomous Intelligent Systems, Technical
report: GMD Report 148, 2001.

\bibitem{key-Jaeger-Haas-Science}H. Jaeger and H. Haas, 
Science \textbf{ 78}, pp. 78-80, 2004

\bibitem{key-MaasNatschlagerMarkram02}W. Maas, T. Natschlager, H.
Markram, 
Neural Computation \textbf{14}, 2531-2560, 2002.

\bibitem{key-Verstraeten-Schrauwen-unification-07}
D. Verstraeten, B. Schrauwen, M. D. Haene, D. Stroobandt, 
Neural Networks \textbf{20}, pp. 391-403, 2007.

\bibitem{key-LukoseviciusJaeger-review}M. Lukocevi\v{c}ius, H.Jaeger,
Computer Science Review 3, pp. 127-149, 2009.

\bibitem{luk} Mantas Luko{\v s}evi{\v c}ius, "A practical guide to applying echo state networks", Neural Networks: Tricks of the Trade, 2, vol. 7700: Springer Berlin Heidelberg, pp. 659-686, 2012.

\bibitem{Pikovsky92}
A. S. Pikovsky, Phys. Lett. A {\bf 165}, pp. 33-36 (1992)

\bibitem{OttGrebogiYorke90}
E. Ott, C. Grebogi, and J. A. Yorke, Phys. Rev. Lett. {\bf 64}, pp. 1196--1199 (1990)  

\bibitem{key-SchrauwenBusingLegenstein2009}B. Schrauwen, L. B{\"u}sing,
and R. Legenstein, 
Advances in Neural Information Processing Systems \textbf{ 21}, 1425. (2009)

\bibitem{key-HermanSchrauwen-infiniteEchoStateNetworks}M. Hermans,
B. Schrauwen, 
 Neural Computation \textbf{24}, pp. 104-133, 2012

\bibitem{key-Williams-inifiniteneuralnetworks}C. K. A. Williams,
Neural Computation \textbf{10}, pp. 1203-1216, (1998)

\bibitem{BuonomanoMaass}D. V. Buonomano and W. Maass, Nature Reviews Neuroscience \textbf{10}, 113-125 (2009)

\bibitem{DerridaPomeau86}B. Derrida and Y. Pomeau, "Random networks of automata: a simple annealed approximation", Europhysics Letters 1, 45 (2007)

\bibitem{Bertschinger04} N. Bertschinger, and T. Natschl{\"a}ger, "Real-time computation at the edge of chaos in recurrent neural networks", Neural Computation 16, 1413-1436 (2004)

\bibitem{key-Paquotetal-optoelectRC}Y. Paquot et al., Scientific Reports \textbf{2}  287, 2012.

\bibitem{key-Larger} L. Larger et al., Optics Express \textbf{20} 3241, 2012.



\end{thebibliography}
\end{document}